\documentclass[11pt]{article}
\usepackage{amssymb}
\usepackage{amsmath}
\usepackage[all]{xy}
\usepackage[dvips]{graphicx}
\numberwithin{equation}{section}

\def\be{\begin{equation}}
\def\ee{\end{equation}}
\def\ba{\begin{align}}
\def\ea{\end{align}}
\def\beq{\begin{eqnarray}}
\def\eeq{\end{eqnarray}}

 \input{epsf}

 \usepackage{epsfig}

\begin{document}

\title{\Large{\bf The $AdS_3$ boundary energy momentum
    tensor, \\ exact in the string length over the curvature radius}}
\author{Jan Troost%$^{a}$
} \date{}
\maketitle
\begin{center}
  % $^{a}$
Laboratoire de Physique Th\'eorique \\
 Unit\'e Mixte du CRNS et
     de l'\'Ecole Normale Sup\'erieure \\ associ\'ee \`a l'Universit\'e Pierre et
     Marie Curie 6 \\ UMR
     8549  %\footnote{Preprint LPTENS-10/XX.} 
\\ \'Ecole Normale Sup\'erieure \\
   $24$ Rue Lhomond Paris $75005$, France
\end{center}

 \begin{abstract}
   We first clarify the relation between boundary perturbations of $AdS_3$ in
   general relativity, and exactly marginal worldsheet vertex operators in
   $AdS_3$ string theory with Neveu-Schwarz Neveu-Schwarz flux.  The
   latter correspond to solutions of the higher derivative low-energy
   tree level effective action to all orders in the string length over the
   curvature radius. We then calculate the exact expression of the boundary
   energy momentum tensor including all these higher derivative corrections
   in a purely bosonic string theory.  The bottom-line
   is a canonical shift in the normalization of the boundary
   energy-momentum tensor corresponding to a shift in the curvature
   radius over the string length squared by the dual Coxeter number of
   the $SL(2,\mathbb{R})$ subalgebra of the space-time Virasoro
   algebra. That allows us to derive the value of the Brown-Henneaux
   central charge including all tree level higher derivative
   corrections in bosonic string theory, in a scheme dictated by the
worldsheet conformal field theory.
\end{abstract}

\newpage

\tableofcontents

\section{Introduction}
A quantum theory of gravity on $AdS_3$ with boundary conditions that
allow for black hole configurations, has an asymptotic symmetry
algebra which includes a (left and right copy of the) Virasoro algebra
\cite{Brown:1986nw}. The algebra of diffeomorphism charges is centrally
extended as shown through a Hamiltonian analysis
in classical general relativity \cite{Brown:1986nw}. An
alternative way to compute the central charge is through the
holographic calculation of the Weyl anomaly \cite{Henningson:1998gx}.

String theory in $AdS_3$ is a consistent theory of quantum gravity
that is likely to have a holographic dual. When we allow for black
hole configurations in the bulk, the asymptotic Virasoro algebra is
part of the symmetry algebra.  Indeed, the space-time Virasoro algebra
was explicitly constructed in terms of operators in the worldsheet
conformal field theory in the case of $AdS_3$ backgrounds with
Neveu-Schwarz Neveu-Schwarz flux \cite{Giveon:1998ns}\cite{de
  Boer:1998pp}\cite{Kutasov:1999xu}\cite{Giveon:2001up}, as well as in
the case of  backgrounds with Ramond-Ramond flux \cite{Ashok:2009jw}.
These constructions within the string worldsheet conformal field theory
contain an infinite set of tree level higher curvature
corrections to the classical general relativity analysis.

In this brief note, it is our aim to provide the necessary
calculational details on the construction of the boundary
energy-momentum tensor that takes into account all these higher
derivative corrections in the context of bosonic $AdS_3$ string theory
with pure NSNS flux. This allows us to calculate the exact tree
level value for the Brown-Henneaux central charge within an expansion
scheme which is natural from the point of view of the worldsheet
description of the string theory. Thus, we illustrate the power of an
exact classical solution to string theory to determine all higher
curvature corrections to a physical quantity, in the absence of
supersymmetry.

\section{The boundary energy-momentum tensor}
In the holographic $AdS/CFT$ correspondence
\cite{Maldacena:1997re}\cite{Gubser:1998bc}\cite{Witten:1998qj} the
conserved boundary energy-momentum tensor couples to the massless
graviton in the bulk. In the first subsection we briefly review how we
solve for the bulk deformation that corresponds to introducing a
source term for the boundary energy-momentum tensor in the context of
general relativity. In the second and third subsection, we show how to
include all tree level higher derivative corrections in the context of
string theory on $AdS_3$ with Neveu-Schwarz Neveu-Schwarz flux.

\subsection{Metric perturbations}
We choose a gauge in which the radial components of the graviton at
the boundary are zero, and we solve the bulk equations of motion for
the graviton with boundary condition $\hat{h}_{ab}$ for the spatial
components of the metric.  For a two-dimensional boundary the
perturbative solution to Einstein's equations with negative
cosmological constant is given by
\cite{Liu:1998bu}\cite{Mueck:1998ug}:
\begin{eqnarray}
h_{\mu\nu}(x_0,x^i) &=& \frac{3}{\pi}\int d^2 x' \frac{1}{|x-x'|^{4}} J_{\mu i}(x-x')
J_{\nu j}(x-x') P_{ijab} \hat{h}_{ab}(x'_i),
\end{eqnarray}
where we work in the Poincar\'e coordinates with background metric:
\begin{eqnarray}
ds^2 &=& \frac{1}{x_0^2} (dx_0^2 + dx_i^2)
\end{eqnarray}
and where we used the definitions:
\begin{eqnarray}
P_{ijab} &=& \frac{1}{2} (\delta_{ia} \delta_{jb} + \delta_{ja} \delta_{ib}) - \frac{1}{2} \delta_{ij} \delta_{ab}.
\nonumber \\
J_{\mu \nu} (x) &=& \delta_{\mu \nu} - \frac{2 x_\mu x_\nu}{|x|^2}
\nonumber \\
|x-x'|^2 &=& x_0^2 + |x_i-x_i'|^2.
\end{eqnarray} 
We use the conventions of \cite{Liu:1998bu} in which all the indices
of the tensor $P$ are contracted with
the flat boundary metric.  On the basis of this perturbative solution
and the equality between the renormalized gravity action and the
boundary generating functional of correlation functions, we can
compute the boundary energy-momentum two-point function in the gravity
approximation \cite{Liu:1998bu}\cite{Mueck:1998ug}. 
% We have that
%$h_{00}=0= h_{0i}$ at infinity. The metric perturbation approaches
%$\hat{h}_{ab}$ at the boundary.

It will be useful to express the metric perturbation in terms of
the following coordinates:
\begin{eqnarray}
ds^2 &=& d \phi^2 + e^{2 \phi} d \gamma d \bar{\gamma}.
\end{eqnarray}
When we concentrate on a source-term for the $\gamma
\gamma$ component of the boundary-energy momentum tensor only,
we can write the bulk gravity solution after perturbation as:
\begin{eqnarray}
h_{\bar \gamma \bar \gamma} &=&  \frac{3}{\pi} \int d^2 \gamma 
\frac{e^{-4 \phi}}{|x-\gamma|^8}
\hat{h}_{\bar{\gamma} \bar{\gamma}}
\nonumber \\
h_{\gamma \gamma} &=& \frac{3}{\pi} \int d^2 \gamma 
\frac{(\bar{\gamma}-\bar{x})^4}{|x-\gamma|^8}
\hat{h}_{\bar{\gamma} \bar{\gamma}}
\nonumber \\
h_{\bar{\gamma} \gamma} &=& - \frac{3}{\pi} \int d^2 \gamma 
e^{-2 \phi} (\bar{\gamma}-\bar{x})^2 
\frac{1}{|x-\gamma|^8} \hat{h}_{\bar{\gamma} \bar{\gamma}}
\nonumber \\
h_{\phi \gamma} &=&  \frac{6}{\pi} \int d^2 \gamma
 \frac{e^{-2 \phi} (\bar{x}-\bar{\gamma})^3}{|x-\gamma|^8} \hat{h}_{\bar{\gamma} \bar{\gamma}}
\nonumber \\
h_{\phi \bar{\gamma}} &=&
\frac{6}{\pi} \int d^2 \gamma \frac{e^{-4 \phi}}{|x-\gamma|^8} (\bar \gamma-\bar x)
\nonumber \\
h_{\phi \phi} &=& \frac{12}{\pi}  \int d^2 \gamma
 \frac{e^{- 4 \phi} (\bar{\gamma}-\bar{x})^2}{|x-\gamma|^8}  \hat{h}_{\bar{\gamma} \bar{\gamma}},
\label{components}
\end{eqnarray}
where $\gamma,\bar{\gamma}$ are the complexified boundary coordinates and
$|x-\gamma|^2= e^{- 2\phi} + (x-\gamma)(\bar x -\bar \gamma)$ is a measure
for the distance squared between a point in the bulk and one on the boundary.

\label{Einstein}

\subsection{Embedding in $AdS_3$ string theory}

\label{treestring}

We want to generalize the above result to include all (tree level)
higher derivative terms in the low-energy effective action in the
context of bosonic string theory on $AdS_3$ with Neveu-Schwarz
Neveu-Schwarz flux.  To that end, we wish to recall the exactly
marginal operator corresponding to the metric deformation reviewed in
the previous subsection. The fact that we deal with a solvable
worldsheet conformal field theory 
is the key to allow for an inclusion of all tree level higher
derivative corrections.

We closely follow the paper \cite{Kutasov:1999xu} and refer to it for
many useful concepts, technical details, as well as our notation.  We
recall that we have a left holomorphic worldsheet current algebra
$J^a$ as well as right anti-holomorphic current algebra $\bar{J}^a$,
which can be conveniently packaged in $x$-dependent currents:
\begin{eqnarray}
  J(x;z) &=& e^{-xJ_0^-} J^+(z) e^{x J_0^-}
  =  k ((x-\gamma)^2 e^{2 \phi} \partial \bar{\gamma} + 2 (x-\gamma) \partial \phi
  - \partial \gamma)
  \nonumber \\ 
  \bar{J}(x;z) &=&e^{-\bar{x} \bar{J}_0^-} \bar{J}^+(\bar{z}) e^{\bar{x} \bar{J}_0^-}  = k ((\bar{x}-\bar\gamma)^2 e^{2 \phi} \bar \partial {\gamma} + 2 (\bar x- \bar \gamma) \bar \partial \phi
  - \bar \partial  \bar \gamma).
\label{currents}
\end{eqnarray}
We also introduce the primary vertex operator $\Phi_1$ of worldsheet
conformal dimension zero, and space-time conformal dimension one
(which is also known as the bulk-to-boundary propagator for a massless
scalar field):
\begin{eqnarray}
\Phi_1 &=& \frac{1}{\pi} \frac{1}{(|\gamma-x|^2 e^\phi + e^{-\phi})^2}.
\label{phione}
\end{eqnarray}
The derivative of the bulk deformation computed above in ordinary
gravity with respect to the boundary metric component $\hat{h}_{\bar
  \gamma \bar \gamma}$ gives rise to a certain bulk deformation
which corresponds to the naive boundary energy-momentum tensor.
The metric components that code the bulk deformation, and therefore
the boundary energy-momentum tensor, correspond to the following
worldsheet vertex operator (as can be derived by combining
equations (\ref{components}, \ref{currents}) and (\ref{phione})): 
\begin{eqnarray}
T(x)
%& \approx & 
%\frac{3k}{\pi} (\Phi_1)^4 (- e^{2 \phi} (\bar x -\bar \gamma)^2 \bar \partial \gamma 
%+ 2 \bar \partial \phi (\bar \gamma - \bar x) +\bar{\partial} \bar{\gamma})
%( e^{2 \phi} (\bar x -\bar \gamma)^2 \partial \gamma 
%+ 2 \partial \phi (-\bar \gamma + \bar x) -{\partial} \bar{\gamma}) 
%\nonumber \\
& \approx & \frac{1}{2k} \int d^2 z (J \partial_x^2 \Phi_1 + 3 \partial_x J \partial_x \Phi_1 + 3 \partial_x^2 J \Phi_1) \bar{J}(\bar{x};\bar{z}).
\label{naive}
\end{eqnarray}
The result of our canonical derivation agrees with the proposal of
\cite{Kutasov:1999xu} to which we refer for many more interesting
observations.  In the worldsheet model, the worldsheet vertex operator
is exactly marginal, and it is a physical vertex operator
\cite{Kutasov:1999xu}. Our derivation of the vertex operator in the
context of general relativity does not yet take into account possible
higher derivative corrections to the boundary energy-momentum tensor
$T(x)$, and we still need to accurately define the composite operator
in the quantum theory on the worldsheet.

\subsection{The space-time energy-momentum tensor}
In this subsection, we wish to calculate the correct normalization of
the boundary energy-momentum tensor, in the presence of all higher
derivative terms coded in tree level string theory. An elementary
argument for the need to modify formula (\ref{naive}) runs as
follows. In bosonic string theory, when the worldshseet $SL(2)$
current algebra has level $k$, the worldsheet energy-momentum tensor
has prefactor $1/(k-2)$ (see e.g. \cite{yellowbook} for a review). The conformal
dimensions of primary operators similarly have a $1/(k-2)$
dependence. Therefore, the bulk masses of states will be a function
of $k-2$, and the boundary conformal dimensions as
well. Thus, the space-time energy-momentum tensor is expected to depend on the
level in the combination $k-2$.

That is one way to argue for the proposal for the exact boundary
energy-momentum tensor:
\begin{eqnarray} 
  T(y) &=&
  \frac{1}{2(k-2)} \int d^2 z (\partial_x^2 \Phi_1 J  + 3  \partial_x \Phi_1 \partial_x J + 3 \Phi_1  \partial_x^2 J) \bar{J}(\bar{x};z).\label{Tprop}
\end{eqnarray}
In the above, we mean an exact equality between the (canonically
normalized) boundary-energy momentum operator and the normal ordered
composite operator on the right (where the components of the composite
operator are ordered as indicated). The normal ordering in the
two-dimensional conformal field theory is the one described for
instance in \cite{yellowbook}, which consists of point splitting and then
subtracting
singularities in the OPE evaluated at the location of the second
operator.  We also introduce the operator:
\begin{eqnarray}
C &=& -  \frac{6}{k-2} \int d^2 z 
 (\Phi_1 J \bar{J}) (z,\bar{z})
\end{eqnarray}
and recall \cite{Kutasov:1999xu} that its derivatives with respect to 
either $x$ or $\bar{x}$ decouple from all worldsheet correlation functions
(since it is a total derivative operator without singularities when it encounters 
 physical insertions on the worldsheet).
Using this fact, we can write an equivalent expression for the boundary
energy-momentum tensor:
\begin{eqnarray}
  \nonumber \\
 T(x) &=&  \frac{1}{2(k-2)} \int d^2 w (\partial_x \Phi_1 \partial_x J +2 \Phi_1 \partial_x^2 J ) \bar{J}(\bar{x};z).
\end{eqnarray}
To prove that this is the canonically
normalized boundary energy-momentum tensors to all orders in higher
derivative corrections, i.e. to all orders in a $1/k$ expansion,
it is sufficient to compute the operator product of the boundary energy-momentum
tensor $T$
with itself.

\subsection*{The $TT$ operator product}
An important operator equation in the following calculations is:
\begin{eqnarray}
\partial \Phi_1 &=& \frac{1}{k-2} \partial_x (J \Phi_1),
\label{derivative}
\end{eqnarray}
which arises from the fact that the action of $\partial$ on the
primary operator $\Phi_1$ is equivalent to the action of the worldsheet
Virasoro generator $L_{-1}$ which can be computed via the worldsheet
energy-momentum tensor which is the Sugawara bi-linear in the currents. The
normalization of the energy-momentum tensor is $1/(k-2)$ for a level
$k$ current algebra\footnote{ Our equation differs from
  equation (4.15) in \cite{Kutasov:1999xu} at higher orders in the
  $1/k$ expansion. That leads to a canonically normalized
  space-time energy-momentum tensor. For the superstring, and
  worldsheet supersymmetric sigma-model discussed from section 8
  onwards in \cite{Kutasov:1999xu}, their equation (4.15) becomes exact.
In a purely bosonic context, it is a good semi-classical approximation.}.
Using a similar equation for deriving with respect to
$\bar{x}$, we obtain the result:
\begin{eqnarray}
 \partial_{\bar{x}} T(x) 
&=& \frac{1}{2i} \oint dz ( \partial_x \Phi_1 \partial_x J
+ 2 \Phi_1  \partial_x^2 J ).
\end{eqnarray}
{From} now on, we again  follow \cite{Kutasov:1999xu} closely and
compute the operator product $ \partial_{\bar x} T(x)T(y)$.
Our calculation is an interesting
space-time analogue of the calculation of the shift in the level of the worldsheet
energy-momentum tensor. Since
it is a little intricate,
we produce it here in some detail.  Let's
split up the calculation in several parts.
\subsubsection*{Preparation}
We will make good use of the OPEs\footnote{We use the
  conventions of \cite{Kutasov:1999xu} and refer to that reference for
  further definitions and details.} between the currents and the
primary fields:
\begin{eqnarray}
J(x;z) J(y;z) &\sim & k \frac{(x-y)^2}{(z-w)^2}
+ \frac{1}{z-w} ( (x-y)^2 \partial_y - 2 (y-x)) J(y;w),
\nonumber \\
J(x;z) \Phi_h(y;w) & \sim
& \frac{1}{z-w} \left( (x-y)^2  \partial_y + 2 h (y-x) \right) \Phi_h(y;w)),
\end{eqnarray}
as well as all of their derivatives with respect to both $x$ and
$y$. These operator products code the chiral current algebra as well
as the affine primary nature of the operator $\Phi_h$ of space-time
dimension $h$ \cite{Kutasov:1999xu}.  Note that although the product
$J(x;z)\Phi_h(y;w)$ is regular at $x=y$ \cite{Kutasov:1999xu}, that is
not true for various derivative operators. Composites of derivative
operators therefore do
require normal ordering.  As stated previously,
we take the derivative operators appearing
in the proposed energy-momentum tensor (\ref{Tprop}) to be normal
ordered in the order indicated.
We also recall the regular operator product:
\begin{eqnarray}
\Phi_1(x;z) \Phi_h(y;w) &=& \delta(x-y) \Phi_h(y;w) + O(z-w)
\end{eqnarray}
which was conjectured in \cite{Kutasov:1999xu} and proven in
\cite{Teschner:1999ug} (given our normalization of the 
vertex operators).

\subsubsection*{The space-time primaries}
In order to show that the operator $\Phi_h(x;z)$ is indeed a
space-time primary of dimension $h$, we wish to compute the operator
product expansion $\Phi_h(x;z) \cdot \partial_{\bar y} T(y)$. The
calculation below should be thought of as talking place inside a
space-time correlation function. We compute the contribution due to
the region where the operator $T$ comes close to the operator $\Phi_h$
on the worldsheet. See \cite{Kutasov:1999xu} as well as
\cite{Aharony:2007rq} for a detailed discussion of why this is
sufficient.  We compute the operator product of composite operators
via a point-splitting procedure and denote by $\lim_{:w' \rightarrow
  w:}$ the limit in which we subtract singularities as per the normal
ordering. See \cite{yellowbook} for a pedagogical discussion of this
standard procedure.  We compute:
\begin{eqnarray}
\Phi_h(x;z) \cdot \partial_{\bar y} T(y)
& \sim & \frac{1}{2i} \oint_z dw
\Phi_h(x,z) \cdot ( \partial_y \Phi_1 \partial_y J + 2 \Phi_1  \partial_y^2 J)
\nonumber \\
& \sim &
\frac{1}{2i}  \oint_z dw \frac{1}{w-z} \lim_{:w' \rightarrow w:}
\nonumber \\
& & 
( \partial_y \Phi_1(y;w') (2 (y-x) \partial_x -2h) \Phi_h(x;w)
+ 2 \Phi_1(y;w') 2 \partial_x \Phi_h(x;w) )
\nonumber \\
& \sim &
 \partial_{\bar y} 
( \frac{h}{(x-y)^2} \Phi_h(x;z)
+ \frac{1}{y-x} \partial_x \Phi_h(x;z)).
\end{eqnarray}
The result is consistent with the standard operator product for a space-time
primary of dimension $h$:
\begin{eqnarray}
T(x) \Phi_h(y) & \sim & \frac{h \Phi_h(y)}{(x-y)^2}+ 
\frac{\partial_y \Phi_h(y)}{x-y}.
\end{eqnarray}
Note that this calculation by itself already fixes the normalization of the
space-time energy-momentum tensor.

\subsubsection*{The current  energy-momentum operator product}
The second intermediate result we wish to compute is the operator
product between the current and the (derivative of the) space-time energy-momentum tensor. We find:
\begin{eqnarray}
J(x;z) \cdot \partial_{\bar y} T(y) & \sim&
\frac{1}{2i} \oint_z dw \lim_{:w' \rightarrow w:}
 \partial_y \Phi_1(y;w') \cdot( \frac{2k(y-x)}{(z-w)^2} + \frac{1}{z-w}
((x-y)^2 \partial_y^2 -2)J(y;w)) 
\nonumber \\
& & 
+ \Phi_1(y;w') \cdot ( \frac{4k}{(z-w)^2} +
\frac{2}{z-w} ( -2 \partial_yJ(y;w)+ 2(y-x) \partial_y^2 J(y;w)))
\nonumber \\
& & + \frac{1}{z-w'} ( 4(y-x) \partial_y + (x-y)^2 \partial_y^2
+2) \Phi_1(y;w') \cdot \partial_y J(y;w)
\nonumber \\
& & 
+ \frac{2}{z-w'} ( (x-y)^2 \partial_y + 2 (y-x)) \Phi_1(y;w')) \cdot \partial_y^2 J(y;w).
\nonumber \\
%  & \sim&
% -\pi (  \partial_y \Phi_1(y) ((x-y)^2 \partial_y^2 -2)J(y)(z)
% \nonumber \\
% & & 
% + \Phi_1(y) 2   ( -2 \partial_yJ(y;w')+ 2(y-x) \partial_y^2 J))(z)
% \nonumber \\
% & & + 2k (y-x) \partial_y \partial \Phi_1(y;z) + 4k \partial \Phi_1(y;z)
% \nonumber \\
% & & +  ( 4(y-x) \partial_y + (x-y)^2 \partial_y^2
% +2) \Phi_1(y) \partial_y J(y) (z)
% \nonumber \\
% & &  ( (x-y)^2 \partial_y + 2 (y-x)) \Phi_1(y))  \partial_y^2 J(y)(z)
% \nonumber \\
% & & - 2 (x-y)^2 \partial_y^2 \partial \Phi_1(y;z)
% + 16 (y-x) \partial_y \partial \Phi_1(y;z)
% - 4 \partial \Phi_1
% \nonumber \\
% & & - 4 (x-y)^2 \partial_y^2 \partial \Phi_1 -4 (y-x) \partial_y \partial \Phi_1
% \nonumber \\
& \sim &
\pi \Big(  2 \partial_y \Phi_1 J
     +2 \Phi_1 \partial_y J
    - 4 (y-x) \partial_y \Phi_1 \partial_y J \nonumber \\
& & 
    - 8 (y-x) \Phi_1 \partial_y^2 J
    - 3 (x-y)^2 \partial_y \Phi_1 \partial_y^2 J
    - (x-y)^2 \partial_y^2 \Phi_1 \partial_y J \Big)
\nonumber \\
& & + \pi
\partial \Big( 4 (k+1) \Phi_1+
(2k+8) (y-x) \partial_y \Phi_1
 +2 (x-y)^2 \partial_y^2 \Phi_1
\Big) (y;z).
\end{eqnarray}
Under the global  $SL(2,\mathbb{R})$ charges
(corresponding to a subgroup of the $SO(2,2)$ isometry group of the
$AdS_3$ space-time), the space-time energy-momentum tensor $T(x)$
transforms as a tensor of weight two. That reasoning also determines
the terms in the above expression that are not total derivatives on the
worldsheet, i.e. the first two lines in the final result.
We will use
derivatives of the above operator product expansion in the 
following calculation.
\subsubsection*{The stress-energy tensor operator product}
Finally, we present some details of  the calculation of the operator
product expansion of the (derivative) of the boundary energy-momentum
tensor.  We spit the operator appearing in the energy-momentum tensor
$T(y)$ such that $\Phi_1$ is at $w'$ and $J$ is at $w$, after which we
take the normal ordered limit $\lim_{: w' \rightarrow w:}$.  We
calculate:
\begin{eqnarray}
\partial_{\bar{x}} T(x) T(y)
& \sim & 
\frac{1}{2(k-2)} \int d^2 w \lim_{: w' \rightarrow w:}
\Big(
( \partial_{\bar{x}} \partial_y
( \frac{1}{(x-y)^2} \Phi_1(y;w') + \frac{\partial_y \Phi_1(y;w')}{x-y}) \cdot \partial_y J(y;w) \bar{J}(\bar{y})
\nonumber \\
&   &  +  \partial_{\bar{x}} 
( \frac{1}{(x-y)^2} \Phi_1(y;w') + \frac{\partial_y \Phi_1(y;w')}{x-y}) \cdot 2 \partial_y^2 J(y;w) \bar{J}(\bar{y}) \Big)
\nonumber \\
& & 
+ \frac{2 \pi}{2(k-2)} \int d^2 w \lim_{: w' \rightarrow w:}\Big(
 \partial_y \Phi_1(w') \cdot 
( (-k-4) \partial_x + 2(y-x) \partial_x^2 ) \partial \Phi_1(x) \bar{J}(\bar{y})
\nonumber \\
& & 
+  \Phi_1(w') \cdot 2 \partial_x^2 \partial \Phi_1(x) \bar{J}(\bar{y})
\nonumber \\
& & + \partial_y \Phi_1(w') \cdot 
(2 \partial_x \Phi_1 \partial_x J + 4 \Phi_1 \partial_x^2 J
+ 3 (x-y) \partial_x \Phi_1 \partial_x^2 J + (x-y) \partial_x^2 \Phi_1 \partial_x J) \bar{J}(\bar{y})
\nonumber \\
& & 
+  \Phi_1(w') \cdot 
(-6 \partial_x \Phi_1 \partial_x^2 J -2 \partial_x^2 \Phi_1 \partial_x J) \bar{J}(\bar{y}) \Big).
\end{eqnarray}
After using equation (\ref{derivative}), the calculation reduces to a
careful manipulation of identities for distributions (of the sort
$\partial_x^n (g (x) \delta(x-y)) = g(y) \partial_x^n
\delta(x-y)$). We find (the $\bar x$-derivative of) the end result:
\begin{equation}
T(x)\cdot T(y) \sim \frac{C}{2(x-y)^4} +
 \frac{2T(y)}{(x-y)^2} + \frac{\partial_y T(y)}{(x-y)}\, ,
\end{equation}
where the central charge operator $C$ is equal to:
\begin{eqnarray}
C &=& -  \frac{6}{k-2} \int d^2 w 
  \Phi_1 J \bar{J}.
\end{eqnarray}
The operator product expansion is indeed of the canonical form, thus confirming
the correct normalization of the boundary energy-momentum tensor, to all orders
in the inverse radius (or inverse string tension) expansion.

A similar calculation shows that the central charge operator $C$ is
indeed central (up to an operator vanishing inside string correlation
functions). Finally, the calculation provides us with the exact
central charge operator $C$ with prefactor $1/(k-2)$.

\section{Final remark}
In gravity, one can evaluate the central charge of the asymptotic
Virasoro algebra(s) either by directly computing the algebra of
charges corresponding to the asymptotic symmetry generators and then
evaluating the central charge on the $AdS_3$ background
\cite{Brown:1986nw}\cite{Barnich:2001jy}, by computing the boundary
energy-momentum tensor two-point function, or by computing the Weyl
anomaly holographically \cite{Henningson:1998gx}. There are contexts
in which one can extend these calculations to include higher
derivative corrections very effectively (see e.g. \cite{Kraus:2005vz})
even without knowing all these terms exactly.

In the context of the conformal field theory description of $AdS_3$
string theory, we find a central charge operator $C$ that can take
different values in different states
\cite{Kutasov:1999xu}\cite{Giveon:2001up}.  We can evaluate the vacuum
expectation value of the operator $C$
in the $AdS_3$ vacuum to obtain the central charge
in that state. One can do this by the techniques developed in
\cite{Teschner:1999ug} and applied in \cite{Maldacena:2001km}\footnote{One uses
the Ward identities computed in appendix A of \cite{Maldacena:2001km} and applies them to 
the operator $\Phi_1 J \bar{J} $ inserted in a two-point function of two
unit operators.}.  We deduce the result that the normalized vev of
the operator $\Phi_1 J \bar{J} $ is exactly $-1$.  We must take note
though that, as argued in \cite{de Boer:1998pp}, the leading
contribution of two insertions of the energy momentum tensor inside a
correlation function comes from a disconnected diagram, where we
factor out the energy-momentum two-point function.  In order for the
remaining factor to correspond to a normalized correlation function, we
need the non-normalized two-point function of the energy-momentum
tensor, i.e. the non-normalized vev of the operator $C$.  We propose
to take this into account in naive fashion by introducing a volume
factor as a normalization factor. The volume is contained in the
vacuum amplitude due to an integration over zero-modes, and it is
mildly renormalized due to oscillator modes.  The tree level
contribution also carries a factor of $g_s^{-2}$ where $g_s$ is the
string coupling constant.  Therefore, we expect the normalization
factor (up to numerical factors) to be $g_s^{-2} (k-2)^{3/2} V_{C}
\approx l_s G_{N}^{-1} (k-2)^{3/2}$ where $V_{C}$ is the volume of the
extra compact directions (in string units), $l_s=\sqrt{\alpha'}$ is
the string length, and $G_N$ is the three-dimensional Newton
constant. The factor $(k-2)^{3/2}$ corresponds to the renormalized
volume of $AdS_3$. We take a definition of the three-dimensional Newton
constant $G_N$ that incorporates possible $\alpha'$-corrections to the
compactified volume\footnote{For instance, if we imagine an $AdS_3
  \times S^3 \times T^{18}$ solution to bosonic string theory, we
  would have a derivative corrected volume of $2 \pi^2 (k+2)^{3/2}
  l_s^3$ for the three-sphere which goes into the definition of the
  three-dimensional Newton constant. The assumption on the $AdS_3$
  string theory that we use here and throughout the paper is that the
  worldsheet conformal field theory is factorized. }.  We find
therefore that the central charge operator $C$ will contribute with a
factor of $ -\frac{1}{k-2} (-1) (k-2)^{3/2} l_s / G_N = (k-2)^{1/2}
l_s/G_N^{}$ to tree level amplitudes with two insertions of the
boundary energy-momentum tensor. The overall $k$-independent
coefficient can be fixed by comparing to the semi-classical gravity
limit.  In conclusion, in this scheme associated to the factorized
worldsheet string theory, the exact central charge will be
$c=\frac{3}{2}\sqrt{k-2}\frac {l_s}{G_N^{}}$. That agrees again with a
shift of the semi-classical space-time radius squared $k$ by the dual Coxeter
number of the $SL(2,\mathbb{R})$ subgroup of the space-time Virasoro
algebra.

\section*{Acknowledgements}
I would like to thank Sujay Ashok, Raphael Benichou, Ashoke Sen and
Kostas Skenderis for useful discussions and encouragement. I
acknowledge the grant ANR-09-BLAN-0157-02 for support.

\end{document}